\documentclass[aps,prl,twocolumn,amsfonts,amssymb,amsmath,10pt,showpacs,showkeys,superscriptaddress]{revtex4-1}
\usepackage{graphicx}%
\usepackage{dcolumn}%
\usepackage{bm}%

\usepackage[usenames]{color}
\usepackage{soul} 
\usepackage[normalem]{ulem} 

\newcommand{\be}{\begin{equation}}
\newcommand{\bea}{\begin{eqnarray}}
\newcommand{\ee}{\end{equation}}
\newcommand{\eea}{\end{eqnarray}}

\begin{document}

\title{Intrinsic Photoconductivity of Ultracold Fermions in Optical Lattices}

\author{J.~Heinze}
\thanks{J.~Heinze, J.~S.~Krauser and N.~Fl\"aschner contributed equally to this work.}
\affiliation{Institut f\"ur Laser-Physik, Universit\"at Hamburg, Luruper Chaussee 149, 22761 Hamburg, Germany}

\author{J.~S.~Krauser}
\thanks{J.~Heinze, J.~S.~Krauser and N.~Fl\"aschner contributed equally to this work.}
\affiliation{Institut f\"ur Laser-Physik, Universit\"at Hamburg, Luruper Chaussee 149, 22761 Hamburg, Germany}

\author{N.~Fl\"aschner}
\thanks{J.~Heinze, J.~S.~Krauser and N.~Fl\"aschner contributed equally to this work.}
\affiliation{Institut f\"ur Laser-Physik, Universit\"at Hamburg, Luruper Chaussee 149, 22761 Hamburg, Germany}

\author{B.~Hundt}
\affiliation{Institut f\"ur Laser-Physik, Universit\"at Hamburg, Luruper Chaussee 149, 22761 Hamburg, Germany}

\author{S.~G\"otze}
\affiliation{Institut f\"ur Laser-Physik, Universit\"at Hamburg, Luruper Chaussee 149, 22761 Hamburg, Germany}

\author{A.~P.~Itin}
\affiliation{Institut f\"ur Laser-Physik, Universit\"at Hamburg, Luruper Chaussee 149, 22761 Hamburg, Germany}
\affiliation{Zentrum f\"ur Optische Quantentechnologien, Universit\"at Hamburg, Luruper Chaussee 149, 22761 Hamburg, Germany}
\affiliation{Space Research Institute, Russian Academy of Sciences, Moscow, Russia}

\author{L.~Mathey}
\affiliation{Institut f\"ur Laser-Physik, Universit\"at Hamburg, Luruper Chaussee 149, 22761 Hamburg, Germany}
\affiliation{Zentrum f\"ur Optische Quantentechnologien, Universit\"at Hamburg, Luruper Chaussee 149, 22761 Hamburg, Germany}

\author{K.~Sengstock}
\email[Corresponding author: ]{klaus.sengstock@physnet.uni-hamburg.de}
\affiliation{Institut f\"ur Laser-Physik, Universit\"at Hamburg, Luruper Chaussee 149, 22761 Hamburg, Germany}
\affiliation{Zentrum f\"ur Optische Quantentechnologien, Universit\"at Hamburg, Luruper Chaussee 149, 22761 Hamburg, Germany}

\author{C.~Becker}
\affiliation{Institut f\"ur Laser-Physik, Universit\"at Hamburg, Luruper Chaussee 149, 22761 Hamburg, Germany}
\affiliation{Zentrum f\"ur Optische Quantentechnologien, Universit\"at Hamburg, Luruper Chaussee 149, 22761 Hamburg, Germany}

\begin{abstract}

\preprint{<report number>}

We report on the experimental observation of an analog to a persistent alternating photocurrent in an ultracold gas of fermionic atoms in an optical lattice. The dynamics is induced and sustained by an external harmonic confinement. While particles in the excited band exhibit long-lived oscillations with a momentum dependent frequency a strikingly different behavior is observed for holes in the lowest band. An initial fast collapse is followed by subsequent periodic revivals. Both observations are fully explained by mapping the system onto a nonlinear pendulum.

\end{abstract}

\maketitle

Photoconductivity describes the change of a material's conductivity following an excitation with photons. If the photon energy is resonant with a band transition, electrons are excited from the valence band to the conduction band and an initial insulator becomes conducting \cite{Bube1960}. Today, photoconductivity is widely used in technological applications such as semiconductor photodiodes and photoresistors. It also provides a powerful probe for novel materials, such as graphene \cite{Lee2008}, transistors made from carbon nanotubes \cite{Freitag2007} or semiconductor nanowires \cite{Ahn2005}. To extend the understanding of such complex materials, atomic quantum gases have proven to be powerful model systems. In this context it is desirable to develop and adopt versatile probing methods \cite{Jaksch1998, Lewenstein2007, Bloch2008}. Owing to its excitational structure in several bands, photoconductivity can provide deeper insight into intra- and interband dynamics as well as orbital effects, which gained much interest in recent years. In the field of quantum gases, multiband interactions and dynamics have been experimentally studied mainly with bosonic atoms \cite{Peik1997, Salger2007, Mueller2007, Clement2009, Ernst2010, Sherson2011, Salger2011, Wirth2011, Fabbri2012, Soltan-Panahi2012}, whereas few work has been performed with fermionic atoms \cite{Koehl2005,Heinze2011,Tarruell2012}.

In this letter, we thoroughly study experimentally and theoretically the particle and hole dynamics of fermionic atoms in an optical lattice. Analogous to photoconductivity measurements in solid state physics, we create uncoupled particle and hole excitations using lattice amplitude modulation. The subsequent dynamics in the combined harmonic and periodic potential is reminiscent of a nonlinear pendulum, which is a paradigm for nonlinear dynamics and is used to model many different quantum systems, like ultracold bosons in a double-well potential \cite{Smerzi1997}, spinor Bose-Einstein condensates \cite{Zhang2005}, semiconductor heterostructures \cite{Alekseev2002} and Josephson junctions \cite{Anderson1964}. In an optical lattice, each individual band independently resembles a pendulum with a different nonlinearity, leading to very distinct dynamics. As a direct consequence, the atoms in the excited band undergo pronounced long-lived oscillations with a momentum dependent frequency. In strong contrast, we observe a fast closing of the holes in the lowest energy band followed by periodic rephasings with a slowly decaying revival amplitude. This behavior stems from the stronger nonlinearity in the lowest energy band caused by the smaller band width as compared to the excited band.

\begin{figure*}[t]
  \centering
  \includegraphics[width=18.0cm]{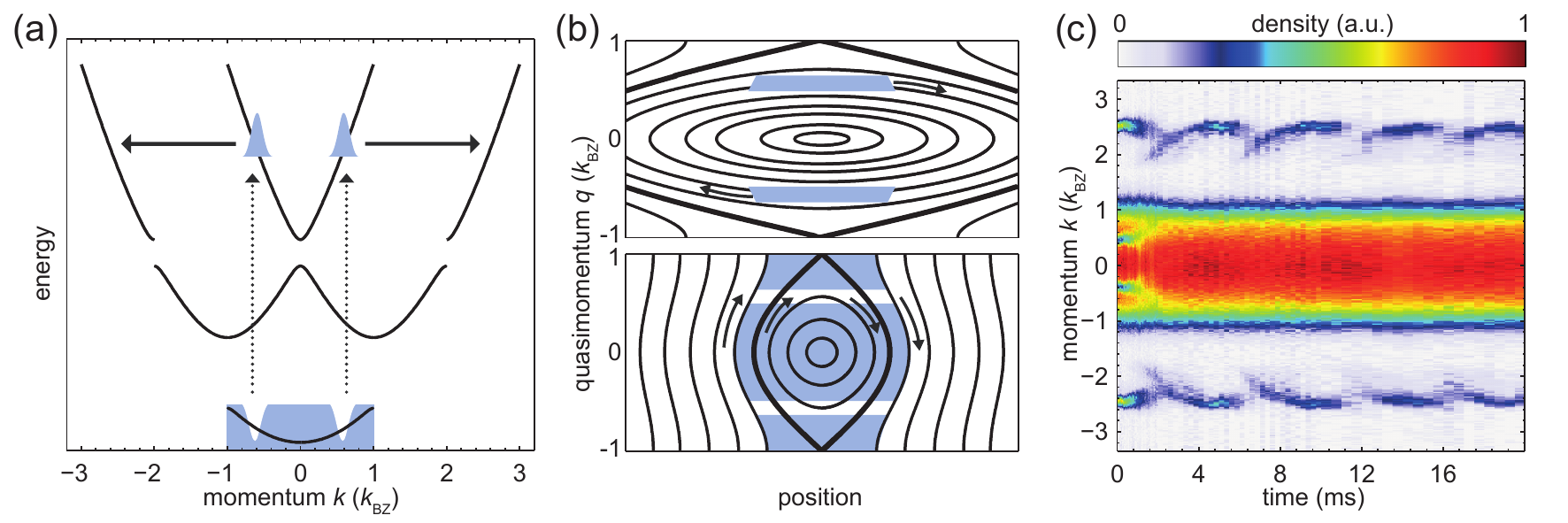}
  \caption{) Principle of particle and hole excitation at a certain quasimomentum $q_0$ via lattice amplitude modulation (dotted lines) and subsequent band mapping process (solid lines) in momentum space. (b) Sketch of the semiclassical phase space of the lowest band (bottom) and second excited band (top). Solid lines depict equal-energy orbits. The shaded area corresponds to the occupied phase space after the lattice amplitude modulation. Due to the different bandwidths, the excited particles occupy only a small region of phase space around a single equal-energy orbit, while the holes are spread over many different orbits. The phase space evolution takes place clockwise along equal-energy orbits (indicated by arrows). (c) Typical photocurrent measurement after excitation to the excited band at $10\,E_\text{r}$. Shown are the column densities of the momentum distribution at different times after the excitation. Atoms in the lowest band are represented by the central plateau. The excitations in the upper band clearly oscillate in momentum space.}
  \label{fig1}
\end{figure*}

In photoconductivity measurements insulators or semiconductors are irradiated with photons which excite electrons to the conduction band at the same time leaving vacancies in the valence band (holes). Both, particle and hole excitations lead to finite conductivity, which can be measured via a photocurrent induced by an external potential. In our system, the photons are mimicked by lattice amplitude modulation, which transfers zero quasimomentum to the system. The frequency of the modulation determines the excitation energy and the initial quasimomentum $q_0$ [see Fig.~\ref{fig1}(a)]. In close analogy to conventional photoconductivity, we create particle (hole) excitations in the second excited (lowest) energy band of the optical lattice, corresponding to the conduction (valence) band. Note, that the created excitations are localized in momentum space in contrast to spatially localized particle-hole excitations in e.g.~Mott insulating systems. Instead of measuring a current through the system, we follow the periodic dynamics of the atoms completely momentum resolved using absorption imaging after time-of-flight. The specific dynamics of particle and hole excitations is induced by an external harmonic confinement, typical for ultracold atom experiments. The total Hamiltonian including harmonic and periodic potential has the form

\begin{equation}
H = \frac{p^2}{2m} + s E_\text{r} \cos(k_\text{BZ} x)^2 + \frac{1}{2} m \omega_0^2 x^2\,,
\label{eq:1}
\end{equation}

with the particle mass $m$, $E_\text{r}=\hbar^2 k_\text{BZ}^2/2m$, $k_\text{BZ}=2\pi/\lambda$, the lattice laser wavelength $\lambda$ and the external trapping frequency $\omega_0$. $s$ determines the lattice depth. The time evolution of fermions in such a potential can be described in a semiclassical phase space \cite{Pezze2004}. In the tight-binding and single-band approximation, the corresponding energy function $H_\text{SC}$ has the form of a nonlinear pendulum, with momentum and position interchanged \cite{Kolovsky2004}

\begin{equation}
H_\text{SC}(x,q) = - 2J \cos(\pi q) + \frac{\nu}{2} x^2\,,
\label{eq:2}
\end{equation}

where $\nu = m \omega_0^2 (\lambda/2)^2$ and $J$ is the tunneling matrix element \cite{Heinze2011}. If the tight-binding approximation is not valid, the momentum dependent part $-2J\cos(\pi q)$ must be replaced by the band dispersion $E^{(n)}_q$. This leads to a nonlinear pendulum with a slightly different potential energy than in (\ref{eq:2}). Figure \ref{fig1}(b) shows the phase space both for the lowest band and the second excited band. Note the existence of a separatrix, where all inner states correspond to closed orbits and all outer states are localized in space \cite{Ott2004}. For the excited band the phase space volume contained within the separatrix is much more extended along the coordinate space direction as compared to the lowest band. This is due to the different band widths given by the individual energy dispersions $E^{(n)}_q$ and leads to the pronounced difference in the behavior of particles in the excited band compared to holes in the lowest band discussed below.

To measure the time evolution of the photocurrent, we prepare either a spin-polarized, non-interacting Fermi gas with $m=9/2$ or an interacting binary spin mixture of $m=-9/2$ and $m=-5/2$ in the $f=9/2$ groundstate manifold of $^{40}$K \cite{SM}. We excite the system via lattice modulation \cite{Heinze2011,Denschlag2002, Stoeferle2004} and detect the quasimomentum distribution by performing adiabatic bandmapping followed by resonant absorption-imaging after typically $15\,\text{ms}$ time-of-flight \cite{Heinze2011,Greiner2001}. Recall, that the bandmapping technique maps particles in the different bands onto their respective Brillouin zones. Our experimental techniques are related to measurements in \cite{Sherson2011} where oscillations of bosonic atoms excited to higher bands were studied. Our method based on a combination of fermionic atoms and detection in momentum space allows to additionally investigate the dynamics of holes and to initialize and detect the dynamics completely momentum resolved.

A typical time evolution in a non-interacting gas is shown in Fig.~\ref{fig1}(c). The atoms in the excited band exhibit a pronounced oscillation in momentum space. Note, that the excitations carry no net current, since always two counterpropagating excitations with quasimomenta $q$ and $-q$ are created due to the inversion symmetry of the band structure. The lifetime of the excitations is on the order of $100\,\text{ms}$, which indicates a recombination between particles and holes that is slow relative to the typical oscillation period. In contrast, for these specific experimental parameters, the holes in the lowest energy band apparently close very fast within the first $2\,\text{ms}$.

\begin{figure}[t]
  \centering
  \includegraphics[width=8.6cm]{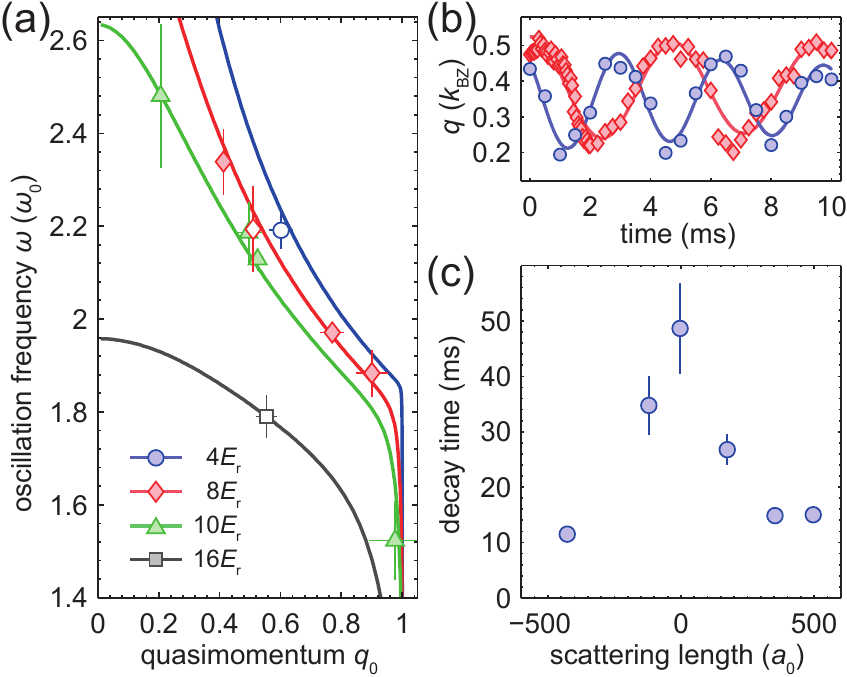}
  \caption{(a) Comparison of measured frequencies with the results of (\ref{eq:3}) for different lattice depths and quasimomenta. Filled symbols represent measurements with non-interacting mixtures. Open symbols show data with interacting mixtures. (b) Oscillations in momentum space of the particles in the excited band at $10\,E_\text{r}$ for two different trapping frequencies $\omega_0=2\pi\times66\,\text{Hz}$ (circles) and $\omega_0=2\pi\times 50\,\text{Hz}$ (diamonds). Solid lines are fits to the data. Extracted oscillation frequencies are $\omega=2\pi\times ( 148\pm5)\,\text{Hz}$ (circles) and $\omega=2\pi\times (107\pm2)\,\text{Hz}$ (diamonds). (c) $1/e$ lifetime of particles in the second excited band as a function of the scattering length for a 3D lattice of $8\,E_\text{r}$. All error bars solely correspond to fit errors, representing two standard deviations.}
  \label{fig3}
\end{figure}

We first concentrate on the excited particles and later comprehensively discuss the hole dynamics. Figures \ref{fig3}(a,b) show the results of photocurrent measurements for a large set of different parameters. Three key observations can be drawn from the experimental data: First, the oscillation frequency $\omega$ linearly depends on the bare trapping frequency $\omega_0$. Second, $\omega$ strongly depends on the initial quasimomentum $q_0$ and third, $\omega$ decreases with increasing lattice depth $s$.

We will now explain all of these observations within the semiclassical nonlinear pendulum picture. As shown in Fig.~\ref{fig1}(b), the excited particles are well localized in the phase space of the upper band and occupy approximately only one single equal-energy orbit. Hence, we model them as a single point in phase space with a given initial quasimomentum $q_0$ and corresponding energy $E(q_0)$, located at the center of the trap. With these initial conditions, the resulting equations of motion can be solved to yield \cite{SM}

\begin{equation}
\omega(q_0) = \omega_0 \frac{\pi}{2}\left(\int_{0}^{q_0}dq\sqrt{\frac{E_\text{r}}{E^{(2)}_{q_0}-E^{(2)}_q}}\right)^{-1}\,.
\label{eq:3}
\end{equation}

The results of (\ref{eq:3}) are shown in Fig.~\ref{fig3}(a) in comparison to the experimental data and show excellent agreement: equation (\ref{eq:3}) directly confirms the linear dependence of $\omega$ on $\omega_0$. More generally, the strong dependence of $\omega$ on $q_0$ is an immediate consequence of the nonlinear pendulum behavior. In particular, for $q_0\rightarrow\pm1$, the excitation approaches the separatrix, where the nonlinear pendulum dynamics is dramatically slowed down and the system eventually reaches a steady state. This is reflected in the strong decrease of the observed frequencies at large $q_0$, visible in Fig.~\ref{fig3}(a). The dependence of $\omega$ on the lattice depth stems from the dependence of the band dispersion $E^{(2)}_q$ on $s$ in (\ref{eq:3}). For small $q_0$, corresponding to small displacements of the pendulum, which lead to a harmonic oscillator behavior, this can be explained in an intuitive picture. In this case, the oscillation frequency depends on the curvature of $E^{(2)}_q$ around $q=0$, which is $d^2E^{(2)}_q/dq^2|_{q=0}$. This curvature decreases with increasing lattice depth leading to a corresponding decrease of $\omega$ as observed in the experiment. To check the validity of the semiclassical approach, we compared the results of (\ref{eq:3}) with a numerical single particle calculation using (\ref{eq:1}) and find perfect agreement, as shown in \cite{SM}.

\begin{figure}[t]
  \centering
  \includegraphics[width=8.6cm]{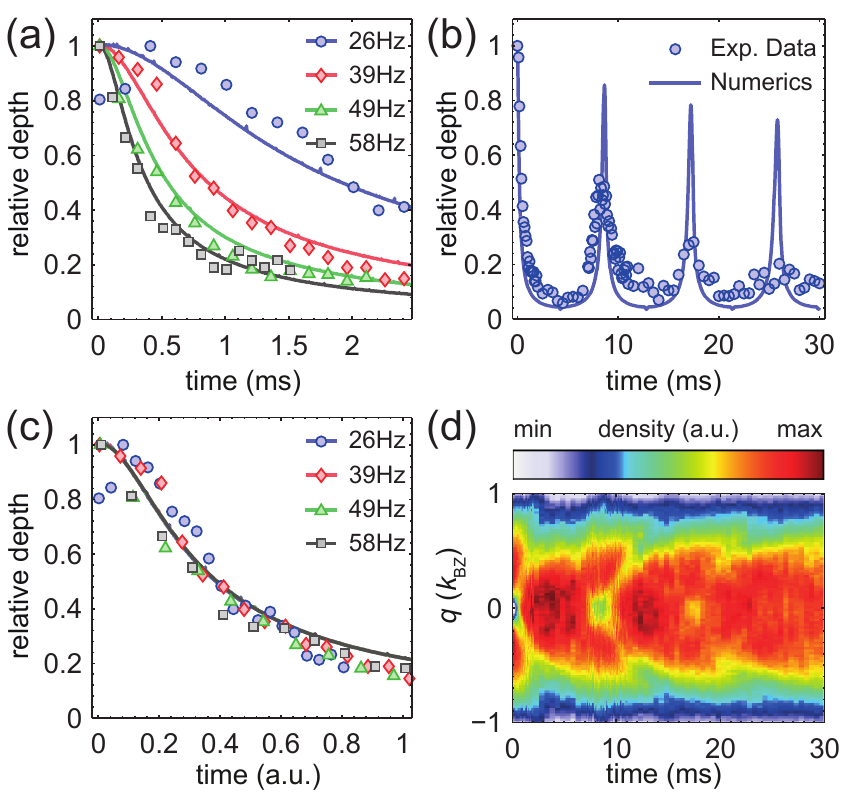}
  \caption{(a) Time evolution of hole depth relative to the maximum depth at $10\,E_\text{r}$ and $q_0=0.5$ for different trapping frequencies $\omega/2\pi$ as given in the legend. Solid lines are semiclassical simulations as described in the text. (b) Rephasing of the hole in quasimomentum for longer evolution times at $2\,E_\text{r}$, $q_0=0.0$ and $\omega_0=2\pi\times63\,\text{Hz}$. Solid line is a semiclassical simulation. (c) Rescaled time evolutions from (a) with scaling factors $\omega_0^2$ as derived from (\ref{eq:4}). The different simulations are not discernible due to the perfect scaling behavior. (d) Momentum resolved data of (b). Only the first Brillouin zone is depicted.}
  \label{fig4}
\end{figure}

We further investigated the influence of interactions on the dynamics and lifetime of the particles in the excited band using a Feshbach resonance at $224\,\text{G}$ \cite{Regal2003}. We observe no effect on the oscillation frequency, whereas we find a substantially reduced lifetime of the atoms in the second excited band for stronger interactions as shown in Fig.~\ref{fig3}(c). Since the total loss of particles is independent of the scattering length as we checked independently, these results can be regarded as a measure for the recombination of particles and holes.

We now return to the hole dynamics in the lowest energy band. In the photoconduction measurement of Fig.~\ref{fig1}(c) we observe a fast reduction of the hole depth within a few ms. This is shown in detail in Fig.~\ref{fig4}(a), for a set of different trapping frequencies. The fast closing cannot be explained by recombination with excited atoms, which have a much longer lifetime, as outlined in the preceding paragraph.

\begin{figure}[t]
  \centering
  \includegraphics[width=8.6cm]{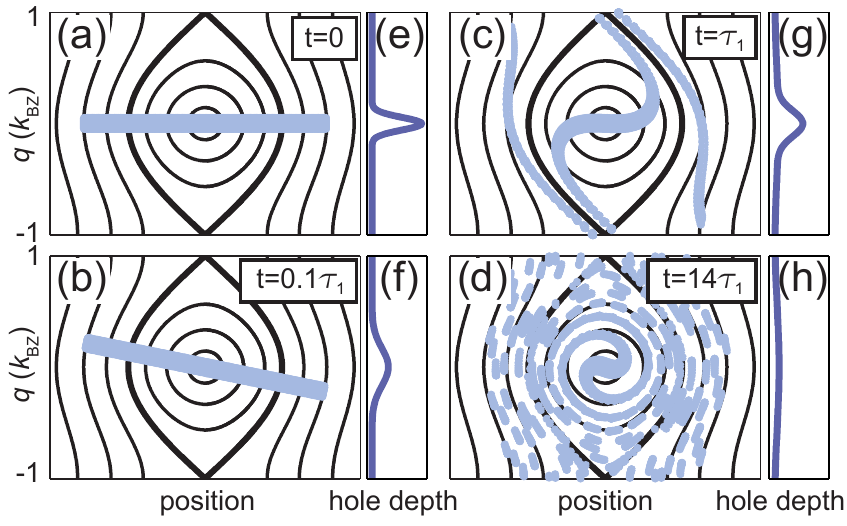}
  \caption{(a)-(d) Sketch of the hole dynamics at $q_0=0$. For short times, the distribution predominantly rotates. For certain parameters a periodic revival of the hole is observed. For long times $t\gg \tau_1$ all trajectories dephase and the hole signature is lost. (e)-(f) Projection of the hole distribution onto momentum space.}
  \label{fig7}
\end{figure}

To describe the holes theoretically, we adopt a description from solid state physics, where holes in a completely filled valence band can be regarded as particles with negative mass \cite{SM}. As sketched in Fig.~\ref{fig7} the phase space distribution of the hole spans over many different equal-energy orbits. This crucial difference as compared to the excited atoms stems from the much smaller bandwidth of the lowest energy band and leads to more complicated dynamics for the holes. Initially, the phase space distribution basically rotates, leading to a fast reduction of the observed hole depth, which is obtained by projecting the phase space distribution onto the quasimonentum axis. For small lattice depth this rotation persists for longer times leading to a revival of the hole in quasimomentum space after multiples of half a rotation in phase space. Indeed, in a measurement specifically addressing this parameter regime, we are able to observe this coherent hole dynamics [Fig.~\ref{fig4}(b,d)]. For strongly nonlinear systems, as realized in deeper lattices, the revival is prevented, since the nonlinearity leads to a deviation from the simple rotation, by introducing momentum dependent oscillation frequencies as in the excited band. Especially for states outside the separatrix this leads to a fast dephasing of the initial phase space distribution resulting in an irreversible disappearance of the holes on our experimental time-scales.

To explore the de- and rephasing of the holes in the nonlinear pendulum picture, we performed simulations on the semiclassical phase space using the Truncated Wigner Approximation method \cite{Polkovnikov2010,Torma2009} for an initial hole distribution which is Gaussian both in momentum and spatial coordinates, with a width $\Delta q$ in momentum space and $\Delta x = 1/\Delta q$ in coordinate space. This corresponds to a coherent superposition of lattice eigenstates as explained in \cite{SM}. Figures \ref{fig4}(a,b) demonstrate the excellent agreement of the simulations with our experimental data. Within the semiclassical description it is also possible to derive an analytical expression for the hole dynamics valid for times $t\ll\tau_1,\tau_2$. $\tau_1=h/\sqrt{2 J\nu}$ is the typical timescale for a full rotation in phase space at small amplitudes. $\tau_2=\hbar \Delta x /\pi J \sin^2(\pi q_0)$ is the timescale for the change of the real space distribution due to the phase space rotation. We find the time dependent depth $D(t)$ of the hole to be \cite{SM}
 
\begin{equation}
D(t) = \frac{ D(t=0)}{\sqrt{1+(t/T)^2}}\,.
\label{eq:4}
\end{equation}

$T=\hbar\pi^2\Delta q^2/\nu$ compares the hole width in momentum space $\Delta q$ to the trap energy, which is a measure for the coupling of different momentum states. Thus, $T$ sets the timescale for changes in the momentum distribution due to rotations in phase space. From (\ref{eq:4}) we find, that the holes are stabilized for broader wavepackets and lower trapping frequencies. In particular, the hole depth scales as $T\propto 1/\omega_0^2$ which allows for stable holes even at moderately low trapping frequencies. Figure \ref{fig4}(c) shows the rescaled simulations and experimental data, confirming the $1/\omega_0^2$ scaling. As for the excited particles, we performed single particle calculations with (\ref{eq:1}) and find very good agreement with the semiclassical description for all parameters \cite{SM}.

In conclusion we have presented a comprehensive study of the dynamics of an excited Fermi gas in an optical lattice, reminiscent of photoconductivity measurements in solids which extends the available techniques to explore dynamical properties of optical lattice systems. We obtain an intuitive and quantitative description of all experimental findings by mapping our system onto the semiclassical phase space of a nonlinear pendulum. In particular, this correctly describes the strikingly different dynamics of particles in the excited band and holes in the lowest energy band. Our results provide the first investigation of holes in momentum space in an ultracold quantum gas in an optical lattice and thus constitute an important contribution towards a more complete understanding of ultracold atoms in periodic potentials. They may also prove crucial for further studies on particle-hole excitations such as excitons \cite{Kantian2007} and gives strong connections to exciting condensed matter systems, which show a similar dynamical behavior \cite{Alekseev2002, Anderson1964}.

We thank P.~T\"orm\"a for valuable discussion. We acknowledge financial support by DFG via Grant FOR801. L.M.~and A.P.I.~acknowledge support from DFG (SFB 925) and the Landesexzellenzinitiative Hamburg, which is financed by the Wissenschaftsstiftung Hamburg and supported by the Joachim Herz Stiftung.

\pagebreak
\onecolumngrid

\section*{Supplemental Information}

\textbf{In the following  supplementary material we start by discussing the preparation of our atomic sample (S1) and give some details of the fitting procedures for the experimental data (S2,S3). We then explain the numerical calculations of quantum dynamics both for particle (S5) and hole (S6) excitations and the semiclassical analysis of particle and hole excitations (S6,S7).}

\section*{S1. Preparation of the atomic sample}

We create a mixture of spin-polarized  $^{40}$K atoms by sympathetic cooling with $^{87}$Rb in a magnetic trap. The atoms are adiabatically transferred to a crossed optical dipole trap operated at $811 \, \text{nm}$ with a $1/e^2$ radius of $120 \, \mu\text{m}$. After switching off the magnetic trap, we remove the rubidium atoms from the trap using a resonant light pulse. For experiments with interacting mixtures, we use a series of rf-pulses and -sweeps to prepare an equal mixture of the hyperfine states $m=-9/2$ and $m=-5/2$ in the hyperfine manifold $f=9/2$. This mixture is then evaporatively cooled by reducing the laser power of the optical dipole trap. We arrive at final particle numbers of about $N = 5-10 \times 10^{4}$ atoms at typical temperatures of $0.2 \, T_{\text{F}}$. After the preparation, we linearly ramp up an optical lattice within $100 \, \text{ms}$. The lattice consists of up to three orthogonal retro-reflected laser beams at $\lambda = 1030 \, {\text{nm}}$ with a $1/e^2$ radius of $200 \, \mu\text{m}$. For measurements at the Feshbach resonance, the magnetic field was set to the final value $50\,\text{ms}$ prior to the $100\,\text{ms}$ optical lattice ramp.

To initialize the photocurrent, we modulate the amplitude of one of the three lattice beams for $1\,\text{ms}$ at a frequency, that is resonant with a transition from the lowest energy band to the second excited band. This excites a fraction of particles and leaves vacancies in the lowest band. Due to the different curvature of the bands, the resonance condition depends on the quasimomentum. By tuning the modulation frequency, we have full control over the quasimomentum of the excited particles. Since lattice amplitude modulation does not imprint any quasimomentum, the holes in the lowest energy band have the same quasimomentum as the particles.

\section*{S2. Analysis of experimental data: Particle excitations}
In this section, we describe how we extract the oscillation frequency of the excited atoms from the experimental data. For each time step, we determine the quasimomentum of the excitation by taking the center-of-mass of the atoms in the excited band independently at positive and negative momentum. To be insensitive to global displacements, we measure the relative distance of the two excitation in momentum space instead of their absolute positions. We extract the oscillation frequencies $\Omega$ from this quantity by fitting an exponentially damped cosine of the form

\begin{equation}
\Delta q(t)=A\exp(-\Gamma t)\cos(\Omega t+\Phi)+ C\,,
\end{equation}

with oscillation amplitude $A$, damping rate $\Gamma$, a phase shift $\Phi$ and a constant offset $C$. In the experiment, always two excitations with opposite initial quasimomenta $q_0$ and $-q_0$ are created. Both excitations independently undergo oscillations in the combined potential of lattice and harmonic trap. After a quarter of an oscillation period, both excitations arrive at $q=0$, are Bragg reflected and continue at the other side of the Brillouin zone, respectively. Since the excitations are not distinguishable, our center-of-mass determination cannot resolve this. Therefore, the extracted data artificially exhibits a turning point of the oscillations at this position. The data thus shows an oscillation with twice the fundamental frequency: $\Omega=2\omega$. This factor of $2$ is corrected in all experimental data.

\section*{S3. Analysis of experimental data: Hole excitations}
\begin{figure}[t]
  \centering
  \includegraphics{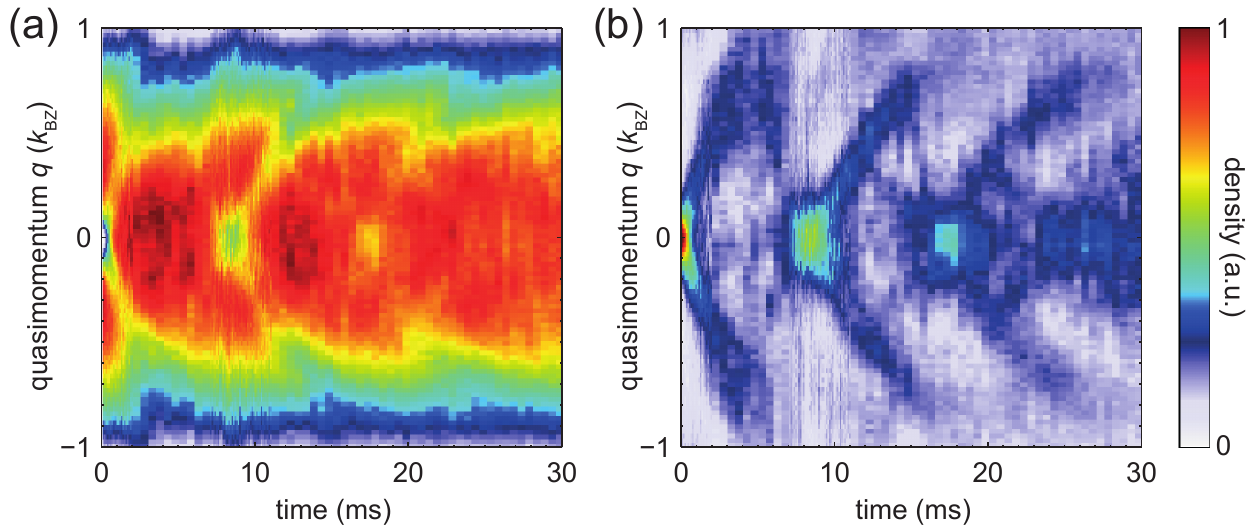}
  \caption{(a) Same data as in Fig.~3(d). (b) Differential image for the data from (a), where the data is subtracted from an offset distribution without lattice amplitude modulation.}
  \label{smfig0}
\end{figure}

To determine the depth of the hole excitation in the lowest energy band we create a differential image by subtracting an independently measured offset picture from all experimental data. For this, we take an absorption picture of the atomic cloud without the lattice amplitude modulation and otherwise identical parameters. The resulting differential image has peaks at the location of the holes and is zero elsewhere. The depth of the hole is taken as the maximum of that differential image. As an example, the differential image of the data from Fig.~3(d) is shown in comparison to the original data in Fig.~\ref{smfig0}.

\section*{S4. Hamiltonian for particle and hole excitations}
In this section, we outline the description of particle and hole excitations in the combined harmonic and periodic potential as generated by an optical lattice. As stated in equation (1) of the manuscript, the Hamiltonian for a single particle in such a potential is

\begin{equation}
H = \frac{p^2}{2m} + s E_\text{r} \cos(k_\text{BZ} x)^2 + \frac{1}{2} m \omega_0^2 x^2\,.
\label{eq:S1}
\end{equation}

By diagonalizing the homogeneous lattice Hamiltonian $H_0=p^2/2m + s E_r \cos(k_\text{BZ} x)^2$, we obtain Bloch states $ \phi_q^{(n)}(x)$ with well defined quasimomentum $q$, band index $n$ and their energy dispersion $E_q^{(n)}$. The effective differential equation for the energy dispersion is $ -\frac{d^2}{dy^2} \phi_q^{(n)}(y) + \frac{s}{4}(2-2\cos(2y))\phi_q^{(n)}(y) = \frac{E_q^{(n)}}{E_\text{r}} \phi_q^{(n)}(y), $ which is equivalent to the Mathieu equation

\be
\frac{d^2 F}{dy^2} + (a+2s_4 \cos(2y))F =0\,, \label{Mathieu}
\ee

for $s_4=\frac{s}{4}$ and $y=k_\text{BZ}x$. The characteristic parameter $a(q,s_4)$ of the Mathieu equation is related to the lattice spectrum like

\be
 a(q,s_4) = \frac{E_q^{(n)}}{E_\text{r}}- 2s_4\,.
\ee

The energy dispersion is characterized by allowed energy bands $E_q^{(n)}$ separated by forbidden band gaps, as shown in Fig.~1(a).
For large enough band gaps, dynamical transfer of atoms between different bands is strongly suppressed and each band can be treated independently. Especially for the lowest energy band, one can safely apply the tight-binding and single-band approximation, which is valid for $s \gtrsim 3$. 
This leads to the discrete Schr\"{o}dinger equation \cite{Kolovsky2004}:

\be i \hbar \dot a_l = \frac{\nu}{2} l^2 a_l - J(a_{l+1}+a_{l-1})\,, \label{DLS} \ee  

with $\nu=m \omega_0^2 (\lambda/2)^2$, and $J$ being the tight-binding tunneling matrix element. $a_l$ is the annihilation operator for a single particle at a given lattice site with index $l$. 
We will use both formulations, the full Hamiltonian in the Bloch basis and the tight-binding, single-band approximation in the following sections.

\section*{S5. Calculation of particle dynamics}
\begin{figure}[t]
  \centering
  \includegraphics{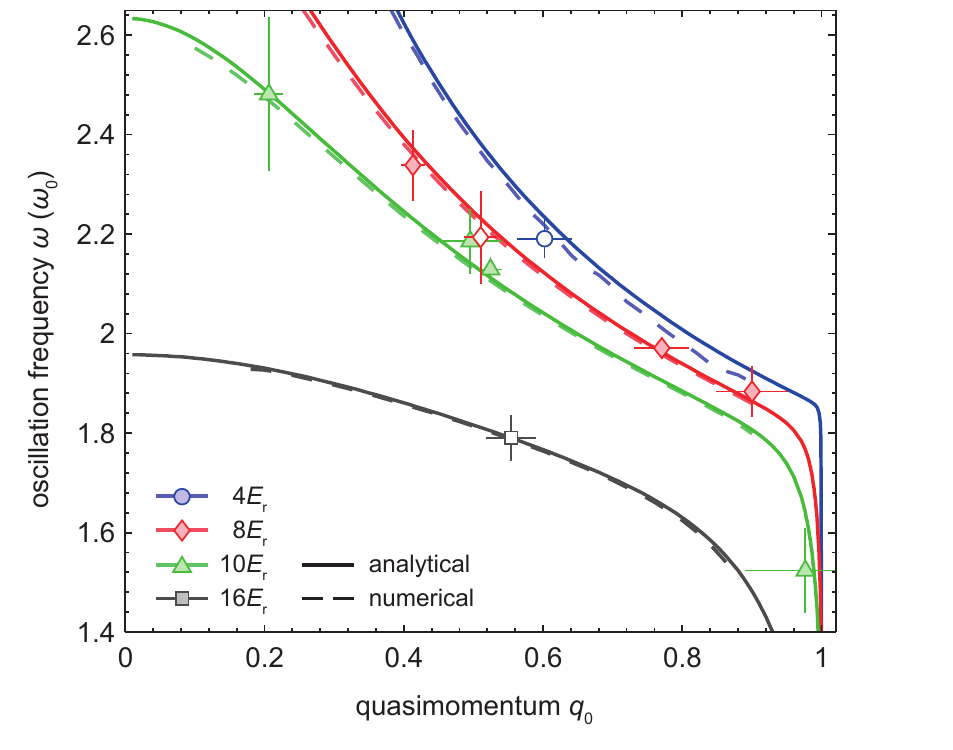}
  \caption{Comparison of measured frequencies, numerical results and semiclassical result for different lattice depth and quasimomenta. Filled symbols represent measurements with non-interacting mixtures. Open symbols show data with interacting mixtures. Solid lines show the results of (\ref{omega}). Dashed lines show numerical calculations using the full trapping and lattice potential.}
  \label{smfig1}
\end{figure}

In this section we outline the numerical description of the particle excitations in the excited band as mentioned in the manuscript.
We assume only a single excited particle, which is in a coherent superposition of multiple Bloch states in the second excited band of the homogeneous lattice Hamiltonian, centered at a given $q_0$. 
This is motivated by the experimental excitation procedure: The excitation is produced by a lattice amplitude modulation pulse of $t=1\,\text{ms}$ duration. 
This leads to a Fourier limited excitation width much larger than the trapping frequency $\omega_0$, such that the harmonic confinement is negligible during the preparation. 
To simplify the calculations, we use a Gaussian distribution of Bloch states with a variance $ \sqrt{v} = \Delta q$ corresponding to an energy of $h \times 364.5\,\text{Hz}$, given by the width of the $1\,\text{ms}$ lattice modulation. 
We obtain the time evolution by exact diagonalization of the full Hamiltonian $H$. 
To extract the oscillation frequencies from the dynamics, we calculate the center-of-mass of the quasi-momentum distribution with respect to the homogeneous lattice Hamiltonian $H_0$ and fit a cosine to the data. 
For a typical calculation we use up to $400$ quasi-momenta and $11$ bands. 
These calculations are compared to the data and the semiclassical results in Fig.~\ref{smfig1} and show excellent agreement.

\section*{S6. Calculation of hole dynamics}
\begin{figure}[t]
  \centering
  \includegraphics{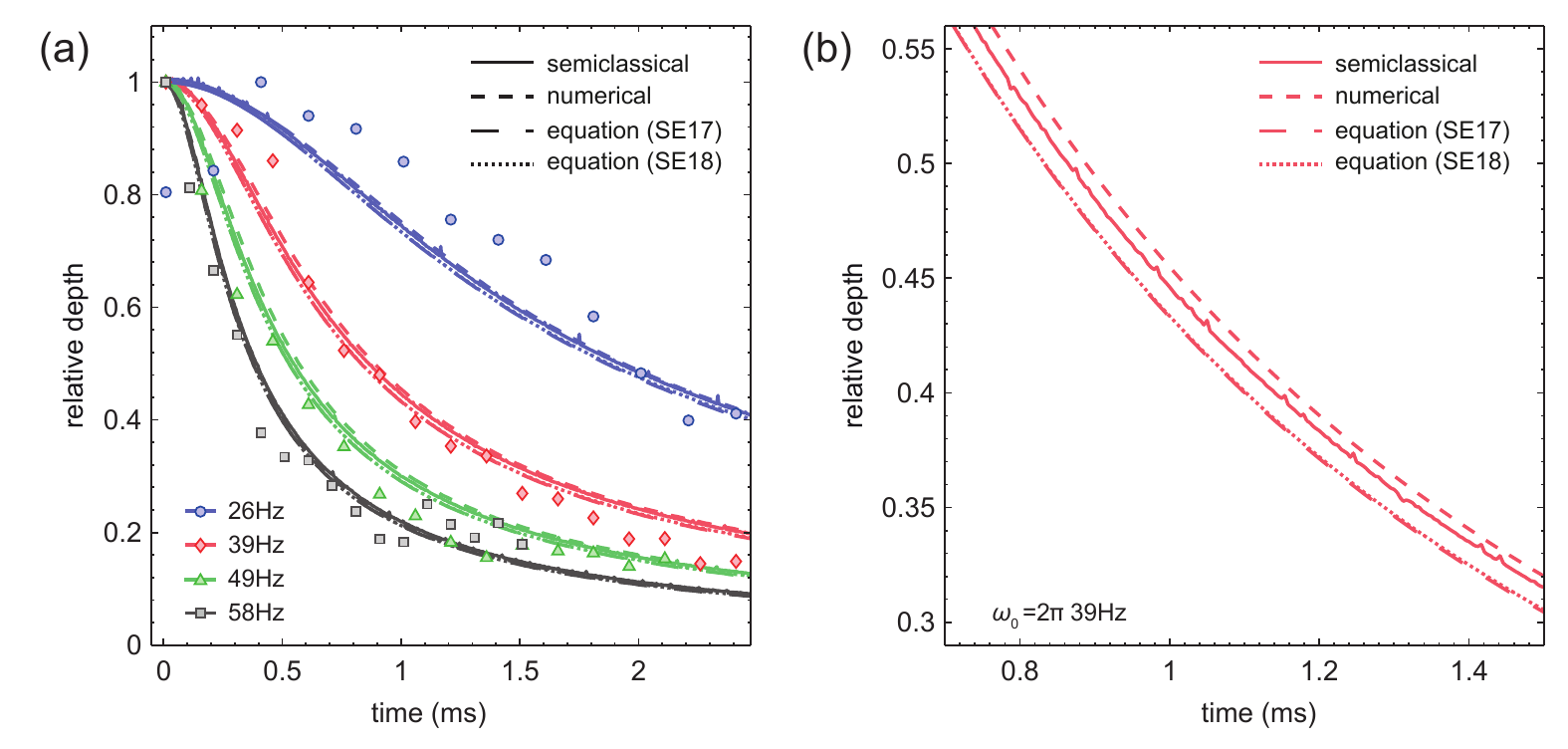}
  \caption{(a) Time evolution of hole depth relative to the maximum depth at $10\,E_\text{r}$ and $q_0=0.5$ for different trapping frequencies $\omega/2\pi$ as given in the legend. Different lines represent the four different calculations as described in the text. (b) Comparison of the four calculations for $\omega_0=2\pi\times 39Hz$ in more detail.}
  \label{smfig2}
\end{figure}

For electron gases in solids a vacancy in an otherwise filled band can be described as a single particle with negative mass. 
Such a hole is in complete analogy to an electron. 
We adopted this description for holes in the lowest band of our harmonically trapped quantum gas. 
To describe this situation numerically, we assumed a single particle with negative mass $m^*=-m$ in the lowest energy band and calculated its evolution in the presence of the harmonic confinement and the lattice potential. 
The excitation has the same shape as in the excited band, since the missing atoms in the lower band directly correspond to excited atoms in the upper band. 
To resemble the experimental situation as closely as possible we also take into account the finite filling of the lowest band. 
This is done by using only the lowest energy states in the time evolution of the initial state. 
To match the occupation number to the experimental situation, we take the total atom number and the trapping frequencies and calculate the number of occupied states in the appropriate direction. 
For comparison with (4) and the semiclassical results, we performed these computations using the tight-binding description of (\ref{DLS}). 
The results are shown in Fig.~\ref{smfig2} in comparison with all other calculations showing a remarkable agreement between all results.

\section*{S7. Semiclassical analysis: effective classical Hamiltonian for the lowest and excited band.}
In the main manuscript we have extensively stressed the close analogy of atoms in a combined periodic and harmonic potential to a non-linear pendulum. In this section we recall the properties of such a pendulum using a phase-space description and explicitly derive the analytic formula for the oscillation frequency.
In a semiclassical description, the Hamiltonian (\ref{eq:S1}) can be mapped onto a semiclassical Hamilton function on a phase space spanned by the dimensionless classical variables $X=x/(\lambda/2)$ and $P=p(\lambda/2)/\hbar=\pi q$. Evaluated in the tight-binding single-band description the semiclassical Hamilton function becomes \cite{Kolovsky2004,Pezze2004}:

\be H(X,P) = - 2J \cos(P) + \frac{\nu}{2} X^2\,. \label{cos} \ee

This is the Hamiltonian of a classical nonlinear pendulum with momentum and coordinate variables interchanged ($X \leftrightarrow P$):
The parabolic confinement plays the role of the kinetic energy of the nonlinear pendulum, while the optical lattice provides an effective (cosine) potential energy of the pendulum. 
As shown in Fig.~1(b), the pendulums equal energy orbits in phase-space can be divided into two categories: 
One is characterized by closed orbits, corresponding to dipole oscillations in the trap or alternatively delocalized states.
The other one is characterized by open lines, corresponding to Bloch oscillations or localized states.
These two types of orbits on  the phase portrait are separated from each other by one characteristic trajectory, the separatrix.
In the case of Bloch oscillations, the wave packet remains on one side of the potential never reaching its center.
For dipole oscillations on the other hand the wave packet passes through the center of the trap repeatedly. 
The frequency of small amplitude oscillations in this system is $\Omega_0=\sqrt{2\nu J}/\hbar$. 
However, the frequency of large-amplitude oscillations strongly depends on the initial amplitude of the oscillation.
With increasing amplitude the frequency decreases and eventually reaches zero at the separatrix.

We now consider the dynamics of a wave packet excited to the second excited band of the lattice, as was done in the experiment. 
In this case, the system exhibits a similar behavior as for the lowest band, but the momentum dependent term $- 2J \cos(P)$ must be replaced by the exact band dispersion $E^{(2)}_P$, deviating from the pure cosine. 
This is possible since all bands can be described independently for large enough band gaps. 
The system still behaves like a nonlinear pendulum, however, with a slightly changed potential energy. 
The corresponding Hamilton function is

\be
H(X,P)= \frac{\nu}{2} X^2 + E_P^{(2)}\,, \label{classical20}
\ee

determining the equations of motion in the usual way. 
Since the excited atoms in the upper band are tightly localized in phase space, they can be regarded as a single point to lowest order. 
Following the experimental parameters, the initial momentum $P_0=\pi q_0$ is tunable, but the excitation is always centered around $X_0=0$ in position space. 
With these initial coordinates, the energy of the excitation is $E_{P_0}^{(2)}$ and position and momentum are related by

\be
X = \sqrt{\frac{2}{\nu}\left(E_{P_0}^{(2)}-E_{P}^{(2)}\right)}\,.
\ee

The oscillations period  $\tau$ for a given initial momentum $P_0$ is given by the integral over the equations of motion. 
We obtain

\be
\frac{\tau}{4}
= \int_{0}^{P_0} \frac{dP}{\nu X}
= \int_{0}^{P_0}  \frac{dP}{ \nu \sqrt{\frac{2}{\nu}(E_{P_0}^{(2)}-E_{P}^{(2)})}} 
= \int_{0}^{P_0} \frac{dP}{\pi \omega_0} \sqrt{\frac{E_R}{(E_{P_0}^{(2)}-E_{P}^{(2)})}}\,,
\ee

with the integral ranging from the center of the Brillouin zone to the initial momentum $P_0=\pi q_0$. 
Substituting $q$ for $P$ as used in the manuscript and using $\omega=2\pi/\tau$ we finally arrive at

\begin{equation}
\omega(q_0) = \omega_0 \frac{\pi}{2}\left(\int_{0}^{q_0}dq\sqrt{\frac{E_\text{r}}{E^{(2)}_{q_0}-E^{(2)}_q}}\right)^{-1}\,,
\label{omega}
\end{equation}

as stated in (3). 
The results of (\ref{omega}) are compared to the experimental data and the numerical calculations in Fig.~\ref{smfig1}. 
All three results show excellent agreement.

\section*{S8. Semiclassical analysis of hole dynamics}

Since the initial phase space distribution of the hole spans over many different equal energy orbits, it is not possible to model the hole as a single point in phase space as done for the particles in the preceding section.
In contrast, to correctly describe the dynamics of the holes it is essential to incorporate the full phase space distribution of the initial state semiclassical wave packet which is an extended object with complicated internal dynamics.

As described in sections (S5,S6) the initial hole wave packet is a coherent superposition of Bloch states with a Gaussian form and a width of $\Delta q$. In the semiclassical description, we need to calculate the Wigner function of the Gaussian wavepacket, which defines the semiclassical phase space distribution. 
We perform the following calculations in the tight-binding and single-band approximation, which is valid in the lowest energy band for our typical experimental parameters. 
For this, a state localized at a single lattice site will be denoted as $|l\rangle$ in the following, where $l$ is the site index.

For a single momentum eigenstate of a lattice without a trap, $|P\rangle = \frac{1}{N} \sum_l e^{i P l} |l \rangle $, the Wigner function
is
\be
W_{l,P}^{(0)} \equiv \sum_{l'} \rho_{l+l',l-l'} e^{-2i P l'}= \frac{1}{N} \sum_{l'} e^{2i(P_0-P)l'}=\delta_{P_0,P}+\frac{1}{N}e^{2i(P_0-P)},
\ee
where $\rho=| P_0\rangle\langle P_0 |$ is the single particle density matrix. For the Gaussian wavepacket, $|\psi \rangle = C \sum_P e^{-\frac{(P-P_0)^2}{2\Delta P^2}}|P\rangle$, the corresponding Wigner function reads

\be
W_{l,P}^G = C' \sum_{l'} e^{2i(P_0-P)l'} e^{-\Delta P^2(l^2+l'^2)} \approx D e^{-\frac{X^2}{1/\Delta P^2}} e^{-\frac{(P_0-P)^2}{\Delta P^2}}\,,
\ee

where $C,C',D$ are normalization constants which are not essential for our discussion. In the last step we replace the summation over site indices with integration, which is justified for a smooth wavepacket. For this $X\cong(\lambda/2)l$ is the continuum limit of the position index $l$.
As a result, the Wigner function of the initial state is Gaussian both in momentum and coordinate with a spatial width of $\Delta X = 1/ \Delta P$ which is inversely proportional to the momentum width. 
Note that the quantum nature of the initial state is effectively encoded in the Wigner function and therefore, in our initial semiclassical distribution. In particular, the Wigner function of a single momentum eigenstate is uniform in coordinate space as shown above, while the Wigner function of the wavepacket is Gaussian in coordinate space due to destructive interference between multiple eigenstates.

To describe the complete hole dynamics, the finite filling of the system due to the trap must be taken into account. 
This corresponds to a projection onto those lowest eigenstates of the lattice and trap potential, which are occupied by  atoms in the lowest band prior to the lattice amplitude modulation. 
For clarity, we present here a zero-temperature analysis, where the lowest energy states are occupied with unity probability up to the Fermi energy and zero probability for $E>E_F$. 
In this case, the finite filling leads to a Wigner function cut-off by a rectangular pulse in coordinate space with ($|X|<X_m$). 
One can estimate the boundary $X_m$ as

\be
X_m = \sqrt{\frac{2(h + 2J \cos (P_0))}{\nu}},
\ee

where $h$ is the energy of the classical Hamiltonian corresponding to the Fermi energy of the quantum system. 
For $P_0=\pi/2$ as used in most of the experiments, this expression simplifies to $X_m=\sqrt{\frac{2h}{\nu}}$. 
For $N$ particles in the lattice, $h$ can be determined from the implicit relation

\be
N = \left\{  \frac{4\sqrt{2}}{\pi}\sqrt{\frac{h+2J}{\nu}} \mbox{E}\left( \frac{4J}{h+2J}\right), \quad \mbox{if} \quad N >N_0,
 \atop \frac{4\sqrt{2}}{\pi}\sqrt{\frac{h+2J}{\nu}} \mbox{E}\left[ \frac{\cos^{-1}\left(\frac{-h}{2J}\right)}{2} ,\frac{4J}{h+2J}\right],     \quad \mbox{if} \quad N<N_0,
 \right.
\ee
where $ N_0 \equiv \frac{8\sqrt{2}}{\pi}\sqrt{\frac{J}{\nu}} $ corresponds to the critical number of particles that fill the entire phase space within the separatrix, $\mbox{E}(.,.)$ is the elliptic integral of the second kind and $\mbox{E}(.)=\mbox{E}(\pi/2,.)$.

For all coherent wavepackets used in the experiment, the cutoff at $X_m$ plays no role, since the Gaussian decays before reaching the boundary. However, $X_m$ might become important for very small particle numbers $N$ or very sharp distributions in momentum space with $\Delta q \lesssim 1/X_m$.

\begin{figure}[t]
  \centering
  \includegraphics[width=0.7\columnwidth]{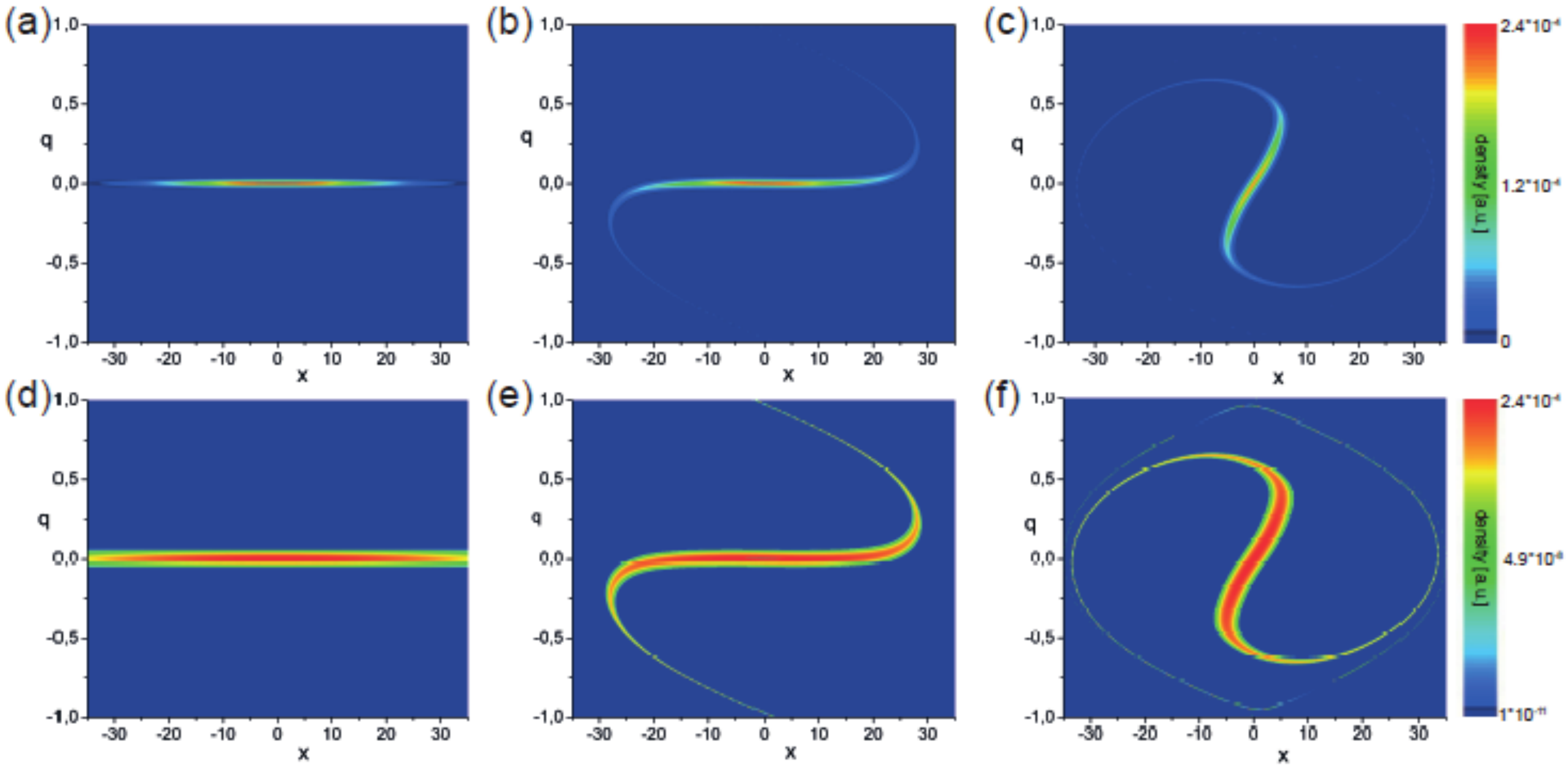}
  \caption{(a) Initial semiclassical distribution of the hole for calculations shown in Fig.~3(b). False color scale is linear. (b) Semiclassical distribution of hole at the first revival after $8.6\,\text{ms}$. (c) Semiclassical distribution after $22.5\,\text{ms}$. (d)-(f) show the same situations as (a)-(c), where the density is shown in a logarithmic scale.}
  \label{smfig3}
\end{figure}

In order to compare the results of the semi-classical approach to the experimental results of Fig.~3, we performed numerical calculations propagating the initial Wigner function of the hole with the semiclassical Hamiltonian equations of motion. 
For this we use $300,000$ phase space points. 
The results are in very good agreement with both the experimental data and the numerical calculations described in (S6) and shown in Fig.~\ref{smfig2}.
Figure \ref{smfig3} shows sample distributions of the calculated time evolution from Fig.~3(b,d). One can nicely observe the swirling within the phases space contained in the seperatrix, leading to a decrease of the hole revival amplitude.

To obtain the analytical expression for the decrease of the relative hole depth as given in (4), we expand the Hamiltonian of the nonlinear pendulum around an initial momentum $P_0$ including  terms up to quadratic order in $P$. 
This approximation holds for times $t\ll\tau_1=2\pi/\Omega_0$. 
We demonstrate the calculation exemplarily for $P_0=\frac{\pi}{2}$, where most of the experiments have been performed.
The resulting linearized equations of motion can be integrated as

\be
P(t) = P_- - \nu X_- t - \frac{1}{2} \Omega_0^2  t^2, \\
\ee

where $P_-$ and $X_-$ are the initial momentum and real space coordinate of a phase point which acquires  the momentum $P(t)$ after a time period $t$. 

We are interested in a situation where a phase point $(X_-,P_-)$ evolves to a point $(X, P_0)$ during time $t$ , since the maximum of the hole at short times is still located at $P=P_0$. 
We need to calculate the relative weight of such trajectories in the whole semiclassical distribution. 
The relative amplitude of the projection of the wavepacket on momentum space at $P=P_0$ is:

\bea
D(t)
 & = & D_*\int dX_- e^{-(\frac{X_-}{\Delta X})^2} e^{-\frac{[\nu X_- t + \frac{1}{2}\Omega_0^2 t^2]^2}{\Delta P^2}}\\
 & = & D(t=0)\frac{1}{\sqrt{1+(t/T)^2}} \exp\left(-\left(\frac{t}{\tau_2}\right)^2\frac{(t/T)^2}{1+(t/T)^2}\right)
\label{D_full}
\eea

with $\tau_2=\hbar \Delta X / J$ and $T=\hbar\Delta P^2/\nu$ and $D_*$ being a normalization constant. 
For times $t$ much smaller than $\tau_2$ this simplifies to:
\be
D(t) \approx  \frac{D(t=0)}{\sqrt{1+(t/T)^2}}\,,\label{D_approx}
\ee
as given in (4) in the manuscript. 
This formula describes the collapse of the hole depth created at $P_0= \pi/2$. 
The general formula for arbitrary $P_0$ contains a modified timescale $\tau_2 =\hbar \Delta X / J \sin^2(P_0)$, but the approximation (\ref{D_approx}) remains valid.

Figure \ref{smfig2} compares all four results for the collapse of the holes: The full numerical treatment, the full semiclassical calculation and the approximate expressions (\ref{D_full}) and (\ref{D_approx}). 
All approaches yield almost identical results.


\begin{thebibliography}{50}

\bibitem{Bube1960}
R.~H.~Bube,
Photoconductivity in solids,
John Wiley and Sons (1960)

\bibitem{Lee2008}
E.~J.~H.~Lee, K.~Balasubramanian, R.~T.~Weitz, M.~Burghard, and K.~Kern,
Nature Nano. \textbf{3}, 486 (2008) 

\bibitem{Freitag2007}
M.~Freitag, J.~C.~Tsang, A.~Bol, D.~Yuan, J.~Liu, and P.~Avouris,
Nano Lett. \textbf{7}, 2037 (2007)

\bibitem{Ahn2005}
Y.~Ahn, J.~Dunning, and J.~Park,
Nano Lett. \textbf{5}, 1367 (2005)

\bibitem{Jaksch1998}
D.~Jaksch, C.~Bruder, J.~I.~Cirac, C.~W.~Gardiner, and P.~Zoller,
Phys. Rev. Lett. \textbf{81}, 3108 (1998)

\bibitem{Lewenstein2007}
M.~Lewenstein, A.~Sanpera, V.~Ahufinger, B.~Damski, A.~Sen and U.~Sen,
Adv. Phys. \textbf{56}, 243 (2007).

\bibitem{Bloch2008}
I.~Bloch, J.~Dalibard, and W.~Zwerger,
Rev. Mod. Phys. \textbf{80}, 885 (2008)

\bibitem{Peik1997}
E.~Peik, M.~B.~Dahan, I.~Bouchoule, Y.~Castin, and C.~Salomon,
Phys. Rev. A \textbf{55}, 2989 (1997)

\bibitem{Salger2007}
T.~Salger, C.~Geckeler, S.~Kling, and M.~Weitz,
Phys. Rev. Lett. \textbf{99}, 190405 (2007)

\bibitem{Mueller2007}
T.~M\"uller, S.~F\"olling, A.~Widera, and I.~Bloch,
Phys. Rev. Lett. \textbf{99}, 200405 (2007) 

\bibitem{Clement2009}
D.~Cl\'ement, N.~Fabbri, L.~Fallani, C.~Fort and M.~Inguscio,
New. J. Phys. \textbf{11}, 103030 (2009)

\bibitem{Ernst2010}
P.~T.~Ernst, S.~G\"otze, J.~S.~Krauser, K.~Pyka, D.-S.~L\"uhmann, D.~Pfannkuche, and K.~Sengstock,
Nature Phys. \textbf{6}, 56 (2010) 

\bibitem{Sherson2011}
J.~F.~Sherson, S.~J.~Park, P.~L.~Pedersen, N.~Winter, M.~Gajdacz, S.~Mai, and J.~Arlt,
New. J. Phys. \textbf{14}, 083013 (2012)

\bibitem{Salger2011}
T.~Salger, C.~Grossert, S.~Kling, and M.~Weitz,
Phys. Rev. Lett. \textbf{107}, 240401 (2011)

\bibitem{Wirth2011}
G.~Wirth, M.~\"Olschl\"ager and A.~Hemmerich,
Nature Phys. \textbf{7}, 147 (2011) 

\bibitem{Fabbri2012}
N.~Fabbri, S.~D.~Huber, D.~Cl\'ement, L.~Fallani, C.~Fort, M.~Inguscio, and E.~Altman,
Phys. Rev. Lett. \textbf{109}, 055301 (2012)

\bibitem{Soltan-Panahi2012}
P.~Soltan-Panahi,	D.-S.~L\"uhmann, J.~Struck,	P.~Windpassinger, and K.~Sengstock,
Nature Phys. \textbf{8}, 71 (2012) 

\bibitem{Koehl2005}
M.~K\"ohl, H.~Moritz, T.~St\"oferle, K.~G\"unter, and T.~Esslinger,
Phys. Rev. Lett. \textbf{94}, 080403 (2005)

\bibitem{Heinze2011}
J.~Heinze, S.~G\"otze, J.~S.~Krauser, B.~Hundt, N.~Fl\"aschner, D.-S.~L\"uhmann, C.~Becker, and K.~Seng\-stock,
Phys. Rev. Lett. \textbf{107}, 135303 (2011)

\bibitem{Tarruell2012}
L.~Tarruell, D.~Greif, T.~Uehlinger, G.~Jotzu, and T.~Esslinger,
Nature \textbf{483}, 302 (2012)

\bibitem{Smerzi1997}
A.~Smerzi, S.~Fantoni, S.~Giovanazzi, and S.~R.~Shenoy,
Phys. Rev. Lett. \textbf{79}, 4950 (1997) 

\bibitem{Zhang2005}
W.~Zhang, D.~L.~Zhou, M.-S.~Chang, M.~S.~Chapman, and L.~You,
Phys. Rev. A \textbf{72}, 013602 (2005) 

\bibitem{Alekseev2002}
K.~N.~Alekseev, and F.~V.~Kusmartsev,
Physics Letters A \textbf{305}, 281 (2002)

\bibitem{Anderson1964}
P.W.~Anderson,
Lectures on the Many-Body Problem (ed. Caianello, E. R.),
Academic, New York (1964)

\bibitem{Pezze2004}
L.~Pezz\`{e}, L. Pitaevskii, A. Smerzi, and S. Stringari, G.~Modugno, E.~de Mirandes, F.~Ferlaino, H.~Ott, G.~Roati, and M.~Inguscio,
Phys. Rev. Lett. \textbf{93}, 120401 (2004)

\bibitem{Kolovsky2004}
A.~R.~Kolovsky, and H.~J.~Korsch,
Int. J. Mod. Phys. B \textbf{18}, 1235 (2004)

\bibitem{Ott2004}
H.~Ott, E.~de Mirandes, F.~Ferlaino, G.~Roati, V.~T\"urck, G.~Modugno, and M.~Inguscio,
Phys. Rev. Lett. \textbf{93}, 120407 (2004)

\bibitem{SM}
For details see supplemental online material.

\bibitem{Denschlag2002}
J.~Hecker~Denschlag, J.~E.~Simsarian, H.~H\"affner, C.~McKenzie, A.~Browaeys, D.~Cho, K.~Helmerson, S.~L.~Rolston, and W.~D.~Phillips,
J. Phys. B: At. Mol. Opt. Phys. \textbf{35}, 3095-3110 (2002)

\bibitem{Stoeferle2004}
T.~St\"oferle, H.~Moritz, C.~Schori, M.~K\"ohl, and T.~Esslinger,
Phys. Rev. Lett. \textbf{92}, 130403 (2004)

\bibitem{Greiner2001}
M.~Greiner, I.~Bloch, O.~Mandel, T.~W.~H\"ansch, and T.~Esslinger,
Phys. Rev. Lett. \textbf{87}, 160405 (2001)

\bibitem{Regal2003}
C.~A.~Regal and D.~S.~Jin,
Phys. Rev. Lett. \textbf{90}, 230404 (2003)

\bibitem{Polkovnikov2010}
A.~Polkovnikov,
Ann.~Phys.~(N.Y.) \textbf{325}, 1790 (2010)

\bibitem{Torma2009}
A.P.~Itin and P.~T\"orm\"a,
Phys. Rev. A \textbf{79}, 055602 (2009)

\bibitem{Kantian2007}
A.~Kantian, A.~J.~Daley, P.~T\"{o}rm\"{a}, and P.~Zoller,
New. J. Phys. \textbf{9}, 407 (2007)

\end{thebibliography}
\end{document}